\documentclass[preprint,showpacs,superscriptaddress,nofootinbib]{revtex4}

\usepackage{amsmath,amssymb,graphicx,color}

\newcommand{\TBM}{{\text{TB}}}
\newcommand{\MNS}{{\text{MNS}}}

\newcommand{\dprime}{{\prime\prime}}
\newcommand{\eV}{{\text{eV}}}

\newcommand{\GeV}{{\text{GeV}}}
\newcommand{\BR}{\text{BR}}
\newcommand{\meee}{\mu \to \bar{e}ee}
\newcommand{\meg}{\mu \to e \gamma}
\newcommand{\llg}{\ell \to \ell^\prime \gamma}
\newcommand{\tlll}{\tau \to \overline{\ell}\ell^\prime \ell^\dprime}

\newcommand{\temm}{\tau \to \overline{e}\mu\mu}
\newcommand{\teLmLmR}{\tau \to \overline{e_L^{}}\mu_L^{} \mu_R^{}}

\newcommand{\tmee}{\tau \to \overline{\mu}ee}
\newcommand{\tmReLeL}{\tau \to \overline{\mu_R^{}} e_L^{} e_L^{}}

\begin{document}
\title{
 Phenomenology in the Zee Model with the $A_4$ Symmetry
}

\author{Takeshi Fukuyama}
\email{fukuyama@se.ritsumei.ac.jp}
\affiliation{Department of Physics and R-GIRO,
Ritsumeikan University, Kusatsu, Shiga 525-8577, Japan}
\affiliation{Maskawa Institute for Science and Culture,
Kyoto Sangyo University, Kyoto 603-8555, Japan}
\author{Hiroaki Sugiyama}
\email{hiroaki@fc.ritsumei.ac.jp}
\affiliation{Department of Physics and R-GIRO,
Ritsumeikan University, Kusatsu, Shiga 525-8577, Japan}
\author{Koji Tsumura}
\email{ko2@phys.ntu.edu.tw}
\affiliation{The Abdus Salam ICTP of UNESCO and IAEA,
Strada Costiera 11, 34151 Trieste, Italy}
\affiliation{Department of Physics, National Taiwan University,
No.~1, Section 4, Roosevelt Road, Taipei, Taiwan}

\begin{abstract}
 The Zee model generates neutrino masses
at the one-loop level by adding
charged $SU(2)_L$-singlet and extra $SU(2)_L$-doublet scalars
to the standard model of particle physics.
 We introduce the softly broken $A_4$ symmetry to the Zee model
as the origin of the nontrivial structure of the lepton flavor mixing.
 This model is compatible with
the tribimaximal mixing
which agrees well with neutrino oscillation measurements.
 Then,
a sum rule
$m_1 e^{i\alpha_{12}^{}} + 2 m_2 + 3 m_3 e^{i\alpha_{32}^{}} = 0$
is obtained
and it results in
$\Delta m^2_{31} < 0$
and
$m_3 \geq 1.8\times 10^{-2}\,\eV$.
 The effective mass $|(M_\nu)_{ee}|$
for the neutrinoless double beta decay
is predicted as
$| (M_\nu)_{ee} | \geq 1.7\times 10^{-2}\,\eV$.
 The characteristic particles in this model
are $SU(2)_L$-singlet charged Higgs bosons
$s^+_\alpha (\alpha=\xi,\eta,\zeta)$
which are made from a ${\bf 3}$ representation of $A_4$.
 Contributions of $s^\pm_\alpha$
to the lepton flavor violating decays of charged leptons
are almost forbidden
by an approximately remaining $Z_3$ symmetry;
 only $\BR(\temm)$ can be sizable
by the flavor changing neutral current interaction
with $SU(2)_L$-doublet scalars.
 Therefore,
$s^\pm_\alpha$ can easily be light enough
to be discovered at the LHC
with satisfying current constraints.
 The flavor structures of $\BR(s^-_\alpha \to \ell \nu)$ are also discussed.
\end{abstract}


\maketitle

\section{introduction}

 The standard model of particle physics~(SM)
can explain almost all of existing experimental results very well.
 However,
the existence of masses of neutrinos,
which are regarded as massless in the SM,
was manifested in 1998 by the evidence
of the atmospheric neutrino oscillation~\cite{Fukuda:1998mi}.
 It is an important question how
the SM should be extended
to generate nonzero neutrino masses.

 For massive neutrinos,
the flavor eigenstates $\nu_{\ell L}^{}\ (\ell=e,\mu,\tau)$,
which are defined by the weak interaction,
are given by superpositions of the mass eigenstates $\nu_{i L}^{}$
as $\nu_{\ell L}^{} = \sum_i (U_\MNS)_{\ell i}\,\nu_{i L}^{}$.
 The mixing matrix $U_\MNS$ is referred to as
the Maki-Nakagawa-Sakata (MNS) matrix~\cite{Maki:1962mu}.
 In the standard parameterization for three neutrinos,
$U_\MNS$ is expressed as
\begin{eqnarray}
U_\MNS
=
 \begin{pmatrix}
  1 & 0 & 0\\
  0 & c_{23} & s_{23}\\
  0 & -s_{23} & c_{23} 
 \end{pmatrix}
 \begin{pmatrix}
  c_{13} & 0 & s_{13}\, e^{-i\delta_D}\\
  0 & 1 & 0\\
  -s_{13}\, e^{i\delta_D} & 0 & c_{13} 
 \end{pmatrix}
 \begin{pmatrix}
  c_{12} & s_{12} & 0\\
  -s_{12} & c_{12} & 0\\
  0 & 0& 1 
 \end{pmatrix} ,
\label{eq:MNS}
\end{eqnarray}
where $s_{ij}$ and $c_{ij}$ denote
$\sin\theta_{ij}$ and $\cos\theta_{ij}$, respectively.
 Brilliant successes of the neutrino oscillation measurements%
~\cite{Fukuda:1998mi,solar,atm,acc,Apollonio:2002gd,:2008ee}
show
\begin{eqnarray}
&&
 \Delta m^2_{21} \simeq 7.6\times 10^{-5}\,\eV^2 , \quad
 |\Delta m^2_{31}| \simeq 2.4\times 10^{-3}\,\eV^2 ,
\label{eq:dm2}\\
&&
 \sin^2{2\theta_{23}} \simeq 1 , \quad
 \sin^2{2\theta_{12}} \simeq 0.87 , \quad
 \sin^2{2\theta_{13}} \lesssim 0.14 ,
\label{eq:theta}
\end{eqnarray}
where $\Delta m^2_{ij} \equiv m_i^2 - m_j^2$,
and $m_i\ (i=1\text{-}3)$ are the mass eigenvalues of $\nu_{iL}^{}$.

 The simplest extension of the SM to accommodate
neutrino masses would be to introduce right-handed neutrinos
which were not necessary for massless neutrinos.
 If neutrino masses are given only by the SM Higgs field
in the same way as the mass generations for charged fermions,
their Dirac masses ($m_D^{} \overline{\nu_L} \nu_R\,+\,\text{h.c.}$)
seem unnaturally smaller than
other fermion masses.
 Then,
we may rely on the seesaw mechanism
where large Majorana mass terms 
for right-handed neutrinos
($M_R \overline{(\nu_R)^c} \nu_R/2\, +\, \text{h.c.}$)
are utilized to obtain very light neutrinos~\cite{seesaw}.
 Such mass terms 
are allowed only for the Majorana particles
which are identical to their antiparticles,
and these terms break the lepton number conservation.

 If neutrino masses are generated
by a completely neutrino-specific mechanism,
any values of neutrino masses seem acceptable
even though they are very different from other fermion masses.
 The Zee model~\cite{Zee:1980ai}
shows an interesting possibility
of such mechanisms.
 In the original Zee model,
an extra $SU(2)_L$-doublet scalar field
and an $SU(2)_L$-singlet charged scalar field
are introduced to the SM\@.
 The mixing between these exotic charged scalars
(from $SU(2)_L$-doublet and singlet fields)
breaks the lepton number conservation.
 Then,
Majorana mass terms of left-handed neutrinos
($m\, \overline{(\nu_L)^c} \nu_L/2\, +\, \text{h.c.}$)
are generated at the one-loop level
without introducing the right-handed neutrino.
 Many works on the model have been done%
~\cite{Zee,Wolfenstein:1980sy,Smirnov:1996bv,Balaji:2001ex,Zee_twozeros,He:2003ih}.

 In the simplest version~\cite{Wolfenstein:1980sy} of the Zee model,
each of the fermions couples with only one of
two $SU(2)_L$-doublet scalars
in order to avoid simply
the flavor-changing neutral current interaction~(FCNC).
 The simplest Zee model was, however, ruled out
at $3\,\sigma$ confidence level~(CL)~\cite{He:2003ih}
by the accumulated knowledge from neutrino oscillation experiments.
 Therefore,
the FCNC should exist in the Zee model as in the original Zee model
with careful consideration about constraints
from the lepton flavor violating (LFV) processes
caused by the FCNC\@.
 It has been shown that
the original Zee model
can satisfy indeed constraints
from neutrino oscillation measurements and
LFV searches~\cite{Balaji:2001ex, He:2003ih}.

 On the other hand,
the nontrivial structure of the lepton flavor mixing
in Eq.~(\ref{eq:theta}) is mysterious
because it is very different from
the simple structure of the quark mixing.
 The lepton sector has
two large mixings
($s_{23}^2 \simeq 0.5$ and
$s_{12}^2 \simeq 0.3$)
while the quark sector has small mixings only.
 It seems natural to expect that
there is some underlying physics
for the special feature of the lepton flavor.
 As a candidate for that,
non-Abelian discrete symmetries
have been studied 
(See, e.g., \cite{discrete} and references therein).
 An interesting choice is the $A_4$ symmetry
because the $A_4$ group is the minimal one
which includes the three-dimensional irreducible representation;
 the representation seems suitable for three flavors of the lepton.
 Some simple models based on the $A_4$ symmetry
can be found in, e.g.,
\cite{A4NR,A4models,A4TB,Ma:2008ym,Ma:2009wi,Fukuyama:2010mz}.
 It is remarkable that
so-called tribimaximal mixing~\cite{Harrison:2002er},
which agrees well with neutrino oscillation data,
can be obtained in an excellent way~\cite{A4TB};
 the tribimaximal mixing is realized
by left-handed lepton doublets in a three-dimensional representation
if the mass eigenstates of charged leptons and neutrinos
are eigenstates of $Z_3$ and $Z_2$ subgroups of $A_4$,
respectively.
 It is very interesting that
the nontrivial mixing structure
is expressed in terms of the symmetry breaking pattern.
 The tribimaximal mixing in the standard parameterization
of Eq.~(\ref{eq:MNS}) is given by
$s_{23}=1/\sqrt{2}$ $(\sin^2{2\theta_{23}}=1)$,
$s_{12}=1/\sqrt{3}$ $(\sin^2{2\theta_{12}}\simeq 0.89)$,
and $s_{13}=0$ as
\begin{eqnarray}
U_\TBM
\equiv
 \begin{pmatrix}
  \sqrt{\frac{2}{\,3\,}}
   & \frac{1}{\sqrt{3}}
   & 0\\
  -\frac{1}{\sqrt{6}}
   & \frac{1}{\sqrt{3}}
   & \frac{1}{\sqrt{2}}\\
  \frac{1}{\sqrt{6}}
   & -\frac{1}{\sqrt{3}}
   & \frac{1}{\sqrt{2}}
 \end{pmatrix} .
\label{eq:TB}
\end{eqnarray}
 Note that models with the $A_4$ symmetry
include several $SU(2)_L$-doublet scalar fields
and each of the leptons couples with more than one of them.
 Therefore,
the lepton sector naturally has the FCNC
which is required for the Zee model to agree
with neutrino oscillation measurements.

 In this article,
we propose the simplest extension of the Zee model
with softly broken $A_4$ symmetry~(A4ZM).
 The soft breaking term of $A_4$ is required
by the appropriate breaking pattern of $A_4$
to obtain the tribimaximal mixing.
 It is assumed that
the soft breaking of $A_4$ is caused by
the small breaking terms of the lepton number conservation
in order to make the phenomenology simple and testable.
 We respect the renormalizability of interactions
and do not introduce
so-called flavons (singlet scalars of the SM gauge group);
 flavons are often introduced to Yukawa interactions
in models with non-Abelian discrete symmetry
as higher dimensional operators
to produce the flavor structure of leptons
(often of quarks also).
 The A4ZM is just for the lepton sector
and the quark sector is almost identical to the SM one.
 Realizing the tribimaximal mixing,
three neutrino masses satisfy a sum rule
$m_1 e^{i\alpha_{12}^{}} + 2 m_2 + 3 m_3 e^{i\alpha_{32}^{}} = 0$.
 Then,
masses and Majorana phases
are governed by only one phase parameter in the A4ZM\@.
 We show $\Delta m^2_{31} < 0$
and lower bounds on $m_3$ and $|(M_\nu)_{ee}|$.
 Although the FCNC is allowed,
we see that most LFV decays
of charged leptons are almost forbidden.
 This is because
a $Z_3$ symmetry remains approximately
and controls well the LFV processes.
 Thus,
it would be expected
that some of the exotic Higgs bosons are light enough
to be discovered at the LHC\@.
 An $SU(2)_L$-singlet charged field $s^+$
is the characteristic particle in the original Zee model,
and the A4ZM includes three fields
$s_\alpha^+$ ($\alpha=\xi, \eta, \zeta$)
which are made from a ${\bf 3}$ representation of $A_4$.
 We discuss
contributions of $s_\alpha^\pm$
to lepton flavor conserving processes;
for example, $\mu \to e \overline{\nu_e} \nu_\mu$,
$\nu_\ell^{} e \to \nu_\ell^{} e$.
 We present
the characteristic flavor structures
of the branching ratios (BRs) of $s_\alpha^- \to \ell \nu$
which will be useful for probing this model at the LHC\@.

 This article is organized as follows.
 In Sec.~\ref{sect:model},
we explain from A to Z of the A4ZM itself.
 It is demonstrated in Sec.~\ref{sect:TB}
how the tribimaximal mixing
can be realized in this model.
 Phenomenology is discussed in Sec.~\ref{sect:pheno}.
 Conclusions and discussions are given in Sec.~\ref{sect:concl}.
 Throughout this article,
we use the words "singlet," etc.\ only for
the representations of $SU(2)_L$
and "${\bf 1}$ representation," etc.\ for
those of $A_4$
in order to avoid confusion.

\section{the model}\label{sect:model}
 The $A_4$ symmetry is characterized by
two elemental transformations $S$ and $T$
which satisfy
\begin{eqnarray}
S^2 = T^3 = (ST)^3 = 1. 
\end{eqnarray}
 There are 3 one-dimensional and 1 three-dimensional
irreducible representations.
 We use the following representations:
\begin{eqnarray}
{\bf 1}&:&
S\, {\bf 1} = {\bf 1}, \quad
T\, {\bf 1} = {\bf 1},\\
{\bf 1}^\prime&:&
S\, {\bf 1}^\prime = {\bf 1}^\prime, \quad
T\, {\bf 1}^\prime = \omega {\bf 1}^\prime,\\
{\bf 1}^\dprime&:&
S\, {\bf 1}^\dprime = {\bf 1}^\dprime, \quad
T\, {\bf 1}^\dprime = \omega^2 {\bf 1}^\dprime,\\
{\bf 3}&:&
S\, {\bf 3}
 = \begin{pmatrix}
    1 &  0 &  0\\
    0 & -1 &  0\\
    0 &  0 & -1
   \end{pmatrix}
   {\bf 3}, \quad
T\, {\bf 3}
 = \begin{pmatrix}
    0 & 1 & 0\\
    0 & 0 & 1\\
    1 & 0 & 0
   \end{pmatrix}
   {\bf 3} ,
\label{eq:sdiag}
\end{eqnarray}
where $\omega \equiv \text{exp}(2\pi i/3)$.
 Since all 12 elements of the $A_4$ group
can be expressed as products of $S$ and $T$,
a model has the $A_4$ symmetry
when the model is invariant under $S$ and $T$.
 For $a = (a_x, a_y, a_z)^T$ and $b = (b_x, b_y, b_z)^T$ of ${\bf 3}$,
the following notations $(ab)_X$
for the decompositions of
${\bf 3}\otimes{\bf 3} \to X$
are used:
\begin{eqnarray}
(a b)_{\bf 1}
&\equiv&
 a_x b_x + a_y b_y + a_z b_z ,\\
(a b)_{{\bf 1}^\prime}
&\equiv&
 a_x b_x + \omega^2 a_y b_y + \omega a_z b_z ,
\\
(a b)_{{\bf 1}^\dprime}
&\equiv&
 a_x b_x + \omega a_y b_y + \omega^2 a_z b_z ,
\\
(a b)_{{\bf 3}_s}
&\equiv&
 \begin{pmatrix}
  a_y b_z + a_z b_y , \
  a_z b_x + a_x b_z , \
  a_x b_y + a_y b_x
 \end{pmatrix}^T ,\\
(a b)_{{\bf 3}_a}
&\equiv&
 \begin{pmatrix}
  a_y b_z - a_z b_y , \
  a_z b_x - a_x b_z , \
  a_x b_y - a_y b_x
 \end{pmatrix}^T .
\end{eqnarray}

\begin{table}[t]
\begin{tabular}{c||c|c|c|c|c|c||c}
 {}
 & $\psi_{1R}^-$
 & $\psi_{2R}^-$
 & $\psi_{3R}^-$
 & $\Psi_{AL}
    = \left(
       \begin{array}{c}
	\psi_{AL}^0\\
	\psi_{AL}^-
       \end{array}
      \right)$
 & $\Phi_A
    = \left(
       \begin{array}{c}
	\phi_A^+\\
	\phi_A^0
       \end{array}
      \right)$
 & \ $s_A^+$ \
 & $\Phi_q
    = \left(
       \begin{array}{c}
	\phi_q^+\\
	\phi_q^0
       \end{array}
      \right)$
\\\hline\hline
 $A_4$
 & ${\bf 1}$
 & ${\bf 1}^\prime$
 & ${\bf 1}^\dprime$
 & ${\bf 3}$
 & ${\bf 3}$
 & ${\bf 3}$
 & ${\bf 1}$
\\\hline
 $SU(2)_L$
 & Singlet
 & Singlet
 & Singlet
 & Doublet
 & Doublet
 & Singlet
 & Doublet
\\\hline
 $U(1)_Y$
 & $-2$
 & $-2$
 & $-2$
 & $-1$
 & $1$
 & $2$
 & $1$
\\\hline
 $L\#$
 & $1$
 & $1$
 & $1$
 & $1$
 & $0$
 & $-2$
 & $0$
\end{tabular}
\caption{
 The leptons and the Higgs bosons in the A4ZM\@.
 The subscript $A = x, y, z$ denotes
the index for ${\bf 3}$ of $A_4$;
 for example,
$(\Psi_{xL}, \Psi_{yL}, \Psi_{zL})$ belongs to ${\bf 3}$
while each $\Psi_{AL}$ are $SU(2)_L$-doublet fields.
 A doublet Higgs field $\Phi_q$ of {\bf 1}
gives masses of quarks
which are assigned to ${\bf 1}$.
 The last row shows assignments of the lepton numbers.
}
\label{tab:particle}
\end{table}

 The particle contents of the A4ZM
are listed in Table~\ref{tab:particle}.
 The excellent realization of the tribimaximal mixing
in models with $A_4$
is achieved by the breaking of $A_4$
into $Z_3$ for charged leptons
and into $Z_2$ for neutrinos~\cite{A4TB}.
 For the appropriate $A_4$ breaking in the charged lepton sector,
left-handed lepton doublets and scalar doublets
should belong to ${\bf 3}$ representations.
 The lepton doublets and scalar doublets
are denoted as $\Psi_{AL}$ and $\Phi_A$, respectively.
 The subscript $A=x,y,z$ stands for
the $A_4$ index of ${\bf 3}$.
 Right-handed charged leptons
$\psi_{1R}^-$, $\psi_{2R}^-$, and $\psi_{3R}^-$
belong to
${\bf 1}$, ${\bf 1}^\prime$, and ${\bf 1}^\dprime$,
respectively%
\footnote{
 See Appendix~\ref{app:R3}
for another case where
right-handed charged leptons
are assigned to a ${\bf 3}$ representation.
}.
 Charged singlet scalars $s^+_A$,
which are the key particles in the Zee model,
are components of a ${\bf 3}$ representation.
 The quark sector is just like the SM
with only particles of ${\bf 1}$.
 The doublet scalar for quark masses
is $\Phi_q$ of ${\bf 1}$.

 The Yukawa terms of leptons with scalar doublets $\Phi_A$
are given~\cite{A4NR} by
\begin{eqnarray}
{\mathcal L}_{\text{d-Yukawa}}
&=&
 y_1 \bigl(\, \overline{\Psi_L}\, \Phi \bigr)_{\bf 1} \psi_{1R}^-
 + y_2 \bigl(\, \overline{\Psi_L}\, \Phi \bigr)_{{\bf 1}^\dprime} \psi_{2R}^-
 + y_3 \bigl(\, \overline{\Psi_L}\, \Phi \bigr)_{{\bf 1}^\prime} \psi_{3R}^-
 + \text{h.c.}
\end{eqnarray}
where $y_i$ are taken to be real
by using redefinitions of phases of $\psi_{iR}^-$.
 The alignment of the vacuum expectation value (vev) is taken as
\begin{eqnarray}
\langle \phi^0_x \rangle
= \langle \phi^0_y \rangle
= \langle \phi^0_z \rangle
= \frac{v_3}{\sqrt{6}} , \quad
\langle \phi^0_q \rangle
= \frac{v_q}{\sqrt{2}} ,
\label{eq:vev}
\end{eqnarray}
where $v_3^2 + v_q^2 = (246\GeV)^2$.
 Note that $v_3^{}$ breaks $A_4$ into $Z_3$ subgroup
while $v_q$ preserves $A_4$.
 The flavor eigenstates of leptons
(the mass eigenstates of charged leptons)
are given by
\begin{eqnarray}
&&
\begin{pmatrix}
 e_R\\
 \mu_R\\
 \tau_R
\end{pmatrix}
=
 \begin{pmatrix}
  \psi_{1R}^-\\
  \psi_{2R}^-\\
  \psi_{3R}^-
 \end{pmatrix}, \\
&&
\begin{pmatrix}
 L_e\\
 L_\mu\\
 L_\tau
\end{pmatrix}
=
 U_L^\dagger
 \begin{pmatrix}
  \Psi_{xL}\\
  \Psi_{yL}\\
  \Psi_{zL}
 \end{pmatrix}, \quad
L_\ell
=
 \begin{pmatrix}
  \nu_{\ell L}\\
  \ell_L
 \end{pmatrix} , \quad
U_L^\dagger
\equiv
 \frac{1}{\sqrt{3}}
 \begin{pmatrix}
  1 &        1 & 1\\
  1 & \omega^2 & \omega\\
  1 &   \omega & \omega^2
 \end{pmatrix} .
\label{eq:lepton}
\end{eqnarray}
 The masses of charged leptons are
\begin{eqnarray}
m_e = \frac{1}{\sqrt{2}}\, v_3^{}\, y_1 , \quad
m_\mu = \frac{1}{\sqrt{2}}\, v_3^{}\, y_2 , \quad
m_\tau = \frac{1}{\sqrt{2}}\, v_3^{}\, y_3 .
\label{eq:clmass}
\end{eqnarray}
 Since $Z_3$ symmetry remains for charged leptons,
each flavor eigenstate has its $Z_3$-charge.
\begin{table}[t]
\begin{tabular}{c||c|c|c}
 {}
  & \ $e_R^{}$, $e_L^{}$, $\nu_{eL}^{}$ \
  & \ $\mu_R^{}$, $\mu_L^{}$, $\nu_{\mu L}^{}$ \
  & \ $\tau_R^{}$, $\tau_L^{}$, $\nu_{\tau L}^{}$
\\
  & \ $s_\xi^+$, $H_{D4}^+$, $H_{D3}^+$ \
  & \ $s_\eta^+$, $H_{D2}^+$ \
  & \ $s_\zeta^+$, $H_{D1}^+$
\\\hline\hline
 \ $Z_3$-charge \
  & $1$
  & $\omega$
  & $\omega^2$
\end{tabular}
\caption{
 List of the $Z_3$-charge
which is conserved approximately in the A4ZM\@.
}
\label{tab:Z3charge}
\end{table}
 Table~\ref{tab:Z3charge} shows $Z_3$-charges of flavor eigenstates.
 The $Z_3$-charges
for $e$, $\mu$, and $\tau$
are $1$, $\omega$, and $\omega^2$, respectively.
 In contrast with the usual flavor quantum number,
$\overline{\mu}$ and $\tau$ have the same $Z_3$-charge $\omega^2$.

 The Yukawa terms with charged singlet scalars $s^\pm_A$
are expressed as
\begin{eqnarray}
{\mathcal L}_{\text{s-Yukawa}}
&=&
 f
  \Bigl(
   \left(
    \overline{(\Psi_L)^c} i\sigma^2 \Psi_L
   \right)_{\bf 3_a}
   s^+
  \Bigr)_{\bf 1}
 + \text{h.c.} ,
\end{eqnarray}
where $f$ is the Yukawa coupling constant,
$\sigma^i (i=1\text{-}3)$ are the Pauli matrices,
and the superscript $c$ means the charge conjugation.
 The antisymmetric nature of the coupling matrix
in the original Zee model ($f_{e\mu} = - f_{\mu e}$, etc.)
is replaced with the antisymmetric decomposition
of two ${\bf 3}$ representations.
 Note that
we can not have such antisymmetric interactions
if $s^\pm$ belongs to one-dimensional representations
(${\bf 1}$, ${\bf 1}^\prime$, and ${\bf 1}^\dprime$).

 In order to obtain the tribimaximal mixing,
the mass eigenstates of neutrinos are required
to be $Z_2$ eigenstates
while flavor eigenstates are $Z_3$ eigenstates.
 Thus,
soft breaking terms of $A_4$ are necessarily introduced to this model.
 It is assumed that
the soft breaking of $A_4$ is
connected to the breaking of the lepton number conservation%
\footnote{
 If the soft breaking of $A_4$ is caused
by the quadratic terms of scalar doublets,
the vacuum in Eq.~(\ref{eq:vev}) will not be natural.
 The quadratic terms of $s_A^\pm$ may be reliable.
 We do not take the option in this article
in order to make phenomenology of $s_A^\pm$ simple.
}.
 Then,
the soft breaking terms are
\begin{eqnarray}
\tilde{V}_\mu
&=&
 \sum_{A=x,y,z}
 \begin{pmatrix}
  \Phi_x^T , \
  \Phi_y^T, \
  \Phi_z^T
 \end{pmatrix}
 \begin{pmatrix}
  0
   & (\mu_A^{})_{xy}
   & (\mu_A^{})_{xz}\\
  -(\mu_A^{})_{xy}
   & 0
   & (\mu_A^{})_{yz}\\
  -(\mu_A^{})_{xz}
   & -(\mu_A^{})_{yz}
   & 0
 \end{pmatrix}
 \begin{pmatrix}
  i\sigma^2 \Phi_x\\
  i\sigma^2 \Phi_y\\
  i\sigma^2 \Phi_z
 \end{pmatrix}
 s^-_A
\nonumber\\
&&\hspace*{20mm}
{}
 +
 \begin{pmatrix}
  \Phi_x^T , \
  \Phi_y^T, \
  \Phi_z^T
 \end{pmatrix}
 \begin{pmatrix}
  (\mu_q^{})_{xx}
   & (\mu_q^{})_{xy}
   & (\mu_q^{})_{xz}\\
  (\mu_q^{})_{yx}
   & (\mu_q^{})_{yy}
   & (\mu_q^{})_{yz}\\
  (\mu_q^{})_{zx}
   & (\mu_q^{})_{zy}
   & (\mu_q^{})_{zz}
 \end{pmatrix}
 \begin{pmatrix}
  s_x^-\\
  s_y^-\\
  s_z^-
 \end{pmatrix}
 i \sigma^2 \Phi_q
 + \text{h.c.}
\label{eq:softA4}
\end{eqnarray}
 Note that
$\tilde{V}_\mu$ does not destroy
the vev alignment (\ref{eq:vev}).
 Since the $\mu$-parameters are the sources
of the neutrino masses,
it seems natural for them to be small%
\footnote
{
 If $\mu$-parameters are large,
mass eigenstates of Higgs bosons are complicated,
and phenomenology on them becomes less predictive.
}.
 The mixings between singlet and doublet scalars become small
because they are controlled by the $\mu$-parameter.
 With the vev's in Eq.~(\ref{eq:vev}),
the soft breaking terms (\ref{eq:softA4})
give the small mixing term as
\begin{eqnarray}
&&
\begin{pmatrix}
 s^-_x, s^-_y, s^-_z
\end{pmatrix}
M^2_{s\phi}
\begin{pmatrix}
 \phi^+_x, \phi^+_y, \phi^+_z, \phi^+_q
\end{pmatrix}^T ,
\\
&&
M^2_{s\phi}
\equiv
 \frac{v_3^{}}{\sqrt{6}}
 \left(
 \begin{matrix}
  2 \left[ (\mu_x^{})_{xy} + (\mu_x^{})_{xz} \right] \
   & 2 \left[ (\mu_x^{})_{yz} - (\mu_x^{})_{xy} \right]\\
  2 \left[ (\mu_y^{})_{xy} + (\mu_y^{})_{xz} \right] \
   & 2 \left[ (\mu_y^{})_{yz} - (\mu_y^{})_{xy} \right]\\
  2 \left[ (\mu_z^{})_{xy} + (\mu_z^{})_{xz} \right] \
   & 2 \left[ (\mu_z^{})_{yz} - (\mu_z^{})_{xy} \right] \
 \end{matrix}
 \right.
\nonumber\\
&&\hspace*{30mm}
 \left.
 \begin{matrix}
  - 2 \left[ (\mu_x^{})_{xz} + (\mu_x^{})_{yz} \right] \
   & - \left[ (\mu_q^{})_{xx} + (\mu_q^{})_{yx} + (\mu_q^{})_{zx} \right]\\
  - 2 \left[ (\mu_y^{})_{xz} + (\mu_y^{})_{yz} \right] \
   & - \left[ (\mu_q^{})_{xy} + (\mu_q^{})_{yy} + (\mu_q^{})_{zy} \right]\\
  - 2 \left[ (\mu_z^{})_{xz} + (\mu_z^{})_{yz} \right] \
   & - \left[ (\mu_q^{})_{xz} + (\mu_q^{})_{yz} + (\mu_q^{})_{zz} \right]
 \end{matrix}
 \right)
\nonumber\\
&&\hspace*{20mm}
{}+
 \frac{v_q}{\sqrt{2}}
 \begin{pmatrix}
  (\mu_q^{})_{xx} \
   & (\mu_q^{})_{yx} \
   & (\mu_q^{})_{zx} \
   & 0\\
  (\mu_q^{})_{xy} \
   & (\mu_q^{})_{yy} \
   & (\mu_q^{})_{zy} \
   & 0\\
  (\mu_q^{})_{xz} \
   & (\mu_q^{})_{yz} \
   & (\mu_q^{})_{zz} \
   & 0
 \end{pmatrix} .
\end{eqnarray}

 Since $\mu$-parameters are assumed to be small,
the $Z_3$ symmetry is preserved approximately also in the Higgs sector.
 Then,
mass eigenstates of scalar fields are
given approximately as eigenstates of $Z_3$.
 For singlet scalars,
mass eigenstates are approximately given by
\begin{eqnarray}
\begin{pmatrix}
 s^+_\xi\\
 s^+_\eta\\
 s^+_\zeta
\end{pmatrix}
=
 U_s^\dagger
 \begin{pmatrix}
  s^+_x\\
  s^+_y\\
  s^+_z
 \end{pmatrix} , \quad
U_s^\dagger
\equiv
 \frac{1}{\sqrt{3}}
 \begin{pmatrix}
  1 & 1 & 1\\
  1 & \omega^2 & \omega\\
  1 & \omega & \omega^2
 \end{pmatrix} .
\end{eqnarray}
 Mass eigenstates of doubletlike charged Higgs bosons $H^+_{Di}$
are expressed with $\tan\beta \equiv v_q/v_3^{}$ approximately as
\begin{eqnarray}
\begin{pmatrix}
 H_{D1}^+\\
 H_{D2}^+\\
 H_{D3}^+\\
 G^+
\end{pmatrix}
=
 U_{\phi^\pm_{}}^\dagger
 \begin{pmatrix}
  \phi_x^+\\
  \phi_y^+\\
  \phi_z^+\\
  \phi_q^+
 \end{pmatrix} , \quad
U_{\phi^\pm_{}}^\dagger
\equiv
 \frac{1}{\sqrt{3}}
 \begin{pmatrix}
  1
   & 0
   & 0
   & 0\\
  0
   & 1
   & 0
   & 0\\
  0
   & 0
   & - s_\beta^{}
   & c_\beta^{}\\
  0
   & 0
   & c_\beta^{}
   & s_\beta^{}
 \end{pmatrix}
 \begin{pmatrix}
  1
   & \omega
   & \omega^2
   & 0\\
  1
   & \omega^2
   & \omega
   & 0\\
  1
   & 1
   & 1
   & 0\\
  0
   & 0
   & 0
   & \sqrt{3}
 \end{pmatrix} ,
\end{eqnarray}
where $c_\beta^{}$ and $s_\beta^{}$ stand for
$\cos\beta = v_3^{}/\sqrt{v_3^2+v_q^2}$ and $\sin\beta$,
respectively.
 Table~\ref{tab:Z3charge} shows $Z_3$-charges of these Higgs bosons also.
 Since a combination $(\Phi_x+\Phi_y+\Phi_z)/\sqrt{3}$
has a $Z_3$-charge $1$,
the combination can be mixed with $\Phi_q$.
 Note that $G^+$ is identified to the Nambu-Goldstone (NG) boson
because neutral partners of $H_{Di}^+$ have no vev.
 The Yukawa coupling constants $(Y_i)_{\ell\ell^\prime}$ for
$(Y_i)_{\ell\ell^\prime}\,
 \overline{\nu_{\ell L}}\, \ell^\prime_R\, H_{Di}^+$
are given by
\begin{eqnarray}
Y_1
=
 \frac{\sqrt{2}}{v_3^{}}
 \begin{pmatrix}
    0 & m_\mu & 0\\
    0 &     0 & m_\tau\\
  m_e &     0 & 0
 \end{pmatrix} , \
Y_2
=
 \frac{\sqrt{2}}{v_3^{}}
 \begin{pmatrix}
    0 &     0 & m_\tau\\
  m_e &     0 & 0\\
    0 & m_\mu & 0
 \end{pmatrix} , \
Y_3
=
 - \frac{\sqrt{2}\,s_{\beta}^{}}{v_3^{}}
 \begin{pmatrix}
  m_e &     0 & 0\\
    0 & m_\mu & 0\\
    0 &     0 & m_\tau
 \end{pmatrix} .
\label{eq:Yi}
\end{eqnarray}
 The matrix of Yukawa coupling constants for $G^+$ is
\begin{eqnarray}
Y_G
=
 \frac{\sqrt{2}\,c_{\beta}^{}}{v_3^{}}
 \begin{pmatrix}
  m_e &     0 & 0\\
    0 & m_\mu & 0\\
    0 &     0 & m_\tau
 \end{pmatrix} .
\label{eq:YG}
\end{eqnarray}
 The Yukawa interactions
$(F_\alpha)_{\ell\ell\prime}\,
 \overline{(\nu_{\ell L}^{})^c}\, \ell^\prime_L\, s_\alpha^+$
are governed by
\begin{eqnarray}
&&
F_\xi
=
 2 i f
 \begin{pmatrix}
  0 & 0 & 0\\
  0 & 0 & -1\\
  0 & 1 & 0
 \end{pmatrix} , \quad
F_\eta
=
 2 i f
 \begin{pmatrix}
  0 & 0 & 1\\
  0 & 0 & 0\\
  -1 & 0 & 0
 \end{pmatrix} , \quad
F_\zeta
=
 2 i f
 \begin{pmatrix}
  0 & -1 & 0\\
  1 & 0 & 0\\
  0 & 0 & 0
 \end{pmatrix} .
\label{eq:Falpha}
\end{eqnarray}
 The small mixing terms of singlet and doublet scalars
are rewritten as
$s_\alpha^-\,
(U_s^\dagger M_{s\phi}^2 U_{\phi^\pm_{}})_{\alpha i}\,
H_{Di}^+$.
 Note that
$s_\alpha^- (U_s^\dagger M_{s\phi}^2 U_{\phi^\pm_{}})_{\alpha 4} G^+ = 0$.

\begin{figure}[t]
\begin{center}
\includegraphics[scale=0.7]{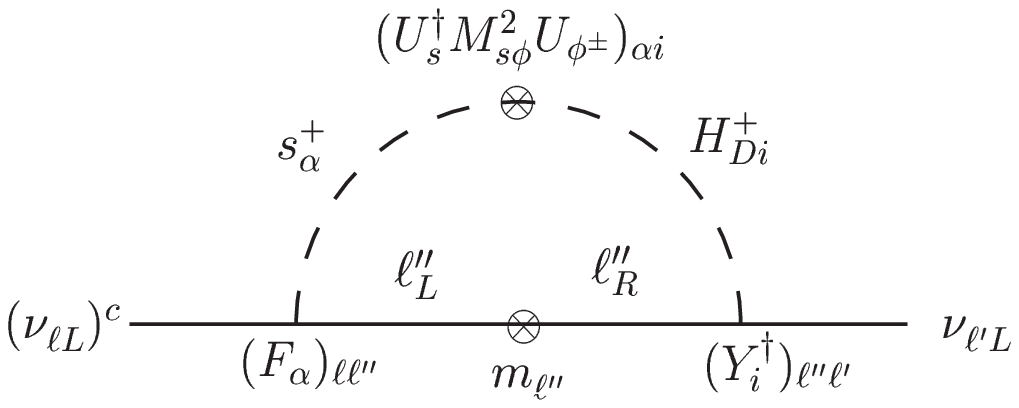}
\includegraphics[scale=0.7]{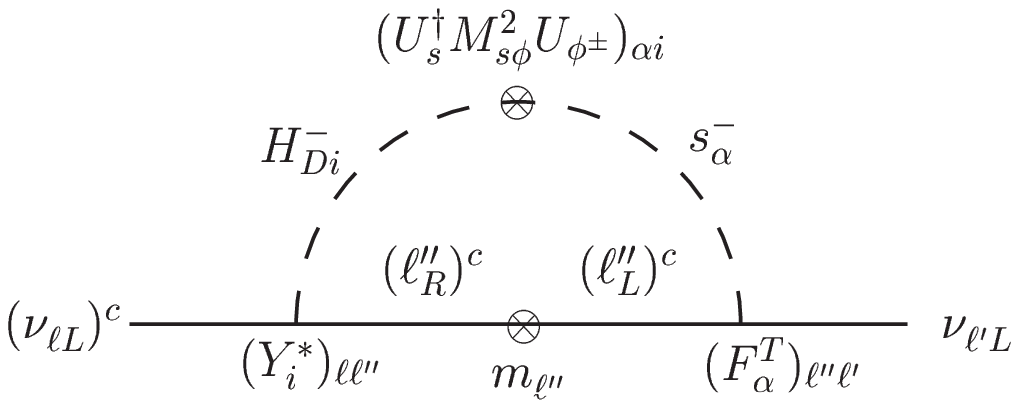}
\vspace*{-5mm}
\caption{
 One-loop diagrams
which generate the mass matrix $M_\nu$
of neutrinos in the flavor basis
in the A4ZM\@.
}
\label{fig:1-loop}
\end{center}
\end{figure}

 The neutrino masses are generated by one-loop diagrams
in the Zee model and also in the A4ZM\@.
 Figure~\ref{fig:1-loop} shows
the one-loop diagrams which generate $M_\nu$
in the flavor basis of neutrinos.
 The expression of $M_\nu$ is
\begin{eqnarray}
(M_\nu)_{\ell\ell^\prime}
&=&
 \sum_{i=1}^3
 \sum_{\ell^\dprime, \alpha}
  (C_{\text{loop}})_{\alpha i}\,
  m_{\ell^\dprime}\,
  (U_s^\dagger M_{s\phi}^2 U_{\phi^\pm_{}})_{\alpha i}
 \Bigl\{
  (F_\alpha)_{\ell \ell^\dprime}
  (Y_i^\dagger)_{\ell^\dprime \ell^\prime}
  +
  (Y_i^\ast)_{\ell \ell^\dprime}
  (F_\alpha^T)_{\ell^\dprime \ell^\prime}
 \Bigr\} ,
\label{eq:Mn0}\\
(C_{\text{loop}})_{\alpha i}
&\equiv&
 -
 \frac{1}{16\pi^2}\,
 \frac{1}{ m_{s^\pm_\alpha}^2 - m_{H_{Di}^\pm}^2 }
 \ln\frac{ m_{s^\pm_\alpha}^2 }{ m_{H_{Di}^\pm}^2 } ,
\end{eqnarray}
where $m_{s^\pm_\alpha}$ and $m_{H_{Di}^\pm}$
are the masses of $s^\pm_\alpha$ and $H_{Di}^\pm$,
respectively.
 There is no contribution from $G^\pm$
because of $(U_s^\dagger M_{s\phi}^2 U_{\phi^\pm_{}})_{\alpha 4} = 0$.

\section{tribimaximal mixing in the A4ZM}\label{sect:TB}
 In this section,
we discuss how the tribimaximal mixing
can be obtained in the A4ZM\@.
 Although other types of mixing can be obtained in this model%
\footnote{
 If the mass matrix $M_{\nu 0}$ for $\psi_{AL}^0$
("neutrinos" in our Lagrangian basis) is diagonalized
by a real $U_\nu$ (orthogonal matrix),
the form of $U_L$ ensures
$|(U_\MNS)_{\mu 3}| = |(U_\MNS)_{\tau 3}|$
which means $\theta_{23}=\pi/4$
in the standard form of $U_\MNS$.
},
models with the $A_4$ symmetry will be motivated well
only when the tribimaximal mixing is achieved
in the leading order approximation.
 The A4ZM gives in general
the following form of
$M_\nu$ in the flavor basis:
\begin{eqnarray}
M_\nu
=
 \frac{m_\tau^2 f}{v_3^{}}
 \begin{pmatrix}
  A_\tau & D_\tau & E_\tau\\
  D_\tau & B_\tau & F_\tau\\
  E_\tau & F_\tau & 0
 \end{pmatrix}
 +
 \frac{m_\mu^2 f}{v_3^{}}
 \begin{pmatrix}
  A_\mu & D_\mu & E_\mu\\
  D_\mu & 0 & F_\mu\\
  E_\mu & F_\mu & C_\mu
 \end{pmatrix}
 +
 \frac{m_e^2 f}{v_3^{}}
 \begin{pmatrix}
  0 & D_e & E_e\\
  D_e & B_e & F_e\\
  E_e & F_e & C_e
 \end{pmatrix} .
\label{eq:formMn}
\end{eqnarray}
 This form of $M_\nu$ is valid also for the original Zee model
where the FCNC is allowed.
 Elements ($A_\tau$, etc.) of the matrix are given by
$\mu$-parameters, vev's,
and $(C_{\text{loop}})_{\alpha i}$ which depends on Higgs boson masses.
 Charged lepton masses appear as squared ones
because of the chirality flip at internal lines
and the forms of $Y_i$ in Eq.~(\ref{eq:Yi}).
 Parts of zeros
are consequence of the antisymmetric nature
of the singlet Yukawa coupling matrices $F_\alpha$.
 The correlation between $m_\ell$ and vanishing elements
is the characteristic feature.
 It is natural that
contributions from $m_e^2$ and $m_\mu^2$ are ignored.
 Then,
we require that $M_\nu$ is diagonalized
by $P U_\TBM$ where
$P\equiv\text{diag}(e^{i\varphi_e}, e^{i\varphi_\mu}, e^{i\varphi_\tau})$
is just a redefinition
of phases of flavor eigenstates
to put the mixing matrix into the standard form.
 The conditions for the diagonalization are
\begin{eqnarray}
B_\tau
&=&
 0 ,
\label{eq:condTB1}\\
E_\tau e^{i\varphi_\tau}
&=&
 - D_\tau e^{i\varphi_\mu} ,
\label{eq:condTB2}\\
F_\tau e^{i(\varphi_\mu+\varphi_\tau)}
&=&
 - A_\tau e^{i\varphi_e}
 - D_\tau e^{i(\varphi_e+\varphi_\mu)} .
\label{eq:condTB3}
\end{eqnarray}
 See \cite{Balaji:2001ex,Zee_twozeros}
for discussions on the original Zee model
with a two-zeros texture
($(M_\nu)_{\mu\mu}=(M_\nu)_{\tau\tau}=0$).
 See also e.g.~\cite{twozeros}
for model-independent discussions
with two-zeros textures.
 With conditions (\ref{eq:condTB1})-(\ref{eq:condTB3}),
the mass eigenvalues can be expressed as
\begin{eqnarray}
m_1 e^{i\alpha_{12}^{}}
&=&
 -|a| + 3 |b| e^{i\varphi} ,
\label{eq:m1}\\
m_2
&=&
 2 |a| ,
\label{eq:m2}\\
m_3 e^{i\alpha_{32}^{}}
&=&
 - |a| - |b| e^{i\varphi} ,
\label{eq:m3}\\
a
&\equiv&
 \frac{m_\tau^2 f}{2v_3^{}}
 \left(
  A_\tau e^{2i \varphi_e}
  + 2 D_\tau e^{i(\varphi_e+\varphi_\mu)}
 \right) ,\\
b
&\equiv&
 \frac{m_\tau^2 f}{2v_3^{}} A_\tau e^{2i \varphi_e} , \quad
\varphi
\equiv
 \text{arg}(b) ,
\end{eqnarray}
where $m_i$ are real and positive.
 Two phases ($\alpha_{12}$ and $\alpha_{31}$)
are the Majorana phases
which are physical parameters
only for Majorana particles~\cite{Mphase}.
 The predictions on neutrinos are discussed
in Sect.~\ref{sect:scale}.
 Ignored masses, $m_e^2$ and $m_\mu^2$
(or Yukawa coupling constants $y_1^2$ and $y_2^2$),
may be regarded as breaking parameters
of the $Z_2$ symmetry in the neutrino sector
which give a deviation from the tribimaximal mixing.
 The deviation will provide a nonzero $\theta_{13}$,
and a naive expectation on the size of $\theta_{13}$ in this model
will be $s_{13} \sim m_\mu^2/m_\tau^2 \simeq 3\times 10^{-3}$.

 In the discussion above,
it was implicitly assumed
that there were sufficient number of parameters
for the neutrino masses
and the tribimaximal mixing.
 An example of undesired situations
is the case where there is no
soft breaking term of $A_4$.
 In this case,
$Z_3$ symmetry remains in the neutrino sector also
and results in $B_\tau=D_\tau=E_\tau=0$
because
$\overline{(\nu_{eL}^{})^c}\,\nu_{\mu L}^{}$,
$\overline{(\nu_{eL}^{})^c}\,\nu_{\tau L}^{}$,
and $\overline{(\nu_{\mu L}^{})^c}\,\nu_{\tau L}^{}$
are forbidden by $Z_3$.
 Then,
$M_\nu$ is constrained too much
to give the tribimaximal mixing
although $\theta_{23}=\pi/4$ can be obtained.
 In addition,
it is impossible to give
a nonzero $\Delta m^2_{32}$.

 Let us demonstrate the realization of the tribimaximal mixing
in the A4ZM in a simple scenario
where $\tan\beta$ is large.
 For example,
the mass ratio of the top quark and the tau lepton,
$m_t/m_\tau \simeq 100$,
seems natural for $\tan\beta$.
 Then,
the $A_4$ symmetry remains approximately in the Higgs sector.
 The mixing between
$\Phi_q$ and $(\Phi_x + \Phi_y + \Phi_z)/\sqrt{3}$
becomes negligible in this case.
 The NG bosons are given dominantly by $\Phi_q$.
 The remaining $A_4$ symmetry gives
almost degenerate masses of exotic Higgs bosons as
$m_{s^\pm_{}} \simeq m_{s^\pm_\alpha}$ and
$m_{\phi^\pm} \simeq m_{H_{Di}^\pm}$.
 The degenerate masses make
the loop function as
an overall factor
$C_{\text{loop}} \simeq (C_{\text{loop}})_{\alpha i}$
of $M_\nu$.
 As a result,
a large $\tan\beta$ simplifies
the conditions (\ref{eq:condTB1})-(\ref{eq:condTB3})
as the ones just between $\mu_q$.
 With $P=(1,1,-1)$ for example%
\footnote
{
 With $P=(1,1,-1)$,
the tribimaximal mixing requires
the mass matrix $M_{\nu 0}$ of $\psi_{AL}^0$
("neutrinos" in our Lagrangian basis)
to satisfy
$(M_{\nu 0})_{xy}=(M_{\nu 0})_{xz}=0$ and
$(M_{\nu 0})_{yy}=(M_{\nu 0})_{zz}$.
 It is clear that $A_4$ is broken to $Z_2$ in the neutrino sector.
},
the conditions (\ref{eq:condTB1})-(\ref{eq:condTB3})
result in
\begin{eqnarray}
(\mu_q)_{zx}
&=&
 \omega (\mu_q)_{xx}
 - \omega^2 (\mu_q)_{yy}
 + (\mu_q)_{zy} ,
\\
(\mu_q)_{yx}
&=&
 \omega^2 (\mu_q)_{xx}
 + (\mu_q)_{yz}
 - \omega (\mu_q)_{zz} ,
\\
(\mu_q)_{xy}
&=&
 \omega^2 (\mu_q)_{zy}
 - \omega^2 (\mu_q)_{xz}
 + (\mu_q)_{yz} .
\end{eqnarray}
 Even in such a simplified case,
$a$ and $b$ are expressed appropriately as
two independent parameters:
\begin{eqnarray}
a
&=&
 \frac{2 \omega f m_\tau^2}{\sqrt{3}}\,
 \frac{v_q}{v_3^{}}\,
 C_{\text{loop}}
 \left\{
  - (\mu_q^{})_{yy}
  + \omega (\mu_q^{})_{zy}
  - (\mu_q^{})_{yz}
  + \omega (\mu_q^{})_{zz}
 \right\} ,\\
b
&=&
 \frac{2 \omega f m_\tau^2}{\sqrt{3}}\,
 \frac{v_q}{v_3^{}}\,
 C_{\text{loop}}
 \left\{
  - (\mu_q^{})_{xz}
  + \omega (\mu_q^{})_{yz}
 \right\} .
\end{eqnarray}
 In the following discussions,
we do not always assume a large $\tan\beta$.
 Strong degeneracy of Higgs boson masses with a large $\tan\beta$
will not be preferred for measuring
characteristic flavor structures of
their leptonic decays.

\section{Phenomenology}\label{sect:pheno}

 Predictions in the A4ZM are discussed in this section.
 Results shown in Sec.~\ref{sect:scale} are
valid not only in the A4ZM
but also in the Zee model (not the simplest one)
with the tribimaximal ansatz.
 Therefore,
in order to test the nature of the $A_4$ symmetry,
decays of Higgs bosons into leptons and LFV decays of charged leptons
are discussed in Secs.~\ref{sect:Hdecay} and \ref{sect:LFV},
respectively.
 Constraints from lepton flavor-conserving decays of charged leptons
are considered in Sec.~\ref{sect:GF}.
 Consequences for neutrino oscillation in matter
are also shown in Sec.~\ref{sect:NSI}.

\subsection{Neutrino masses and Majorana phases}\label{sect:scale}
\begin{figure}[t]
\begin{center}
\includegraphics[scale = 0.3, origin = c, angle = -90]{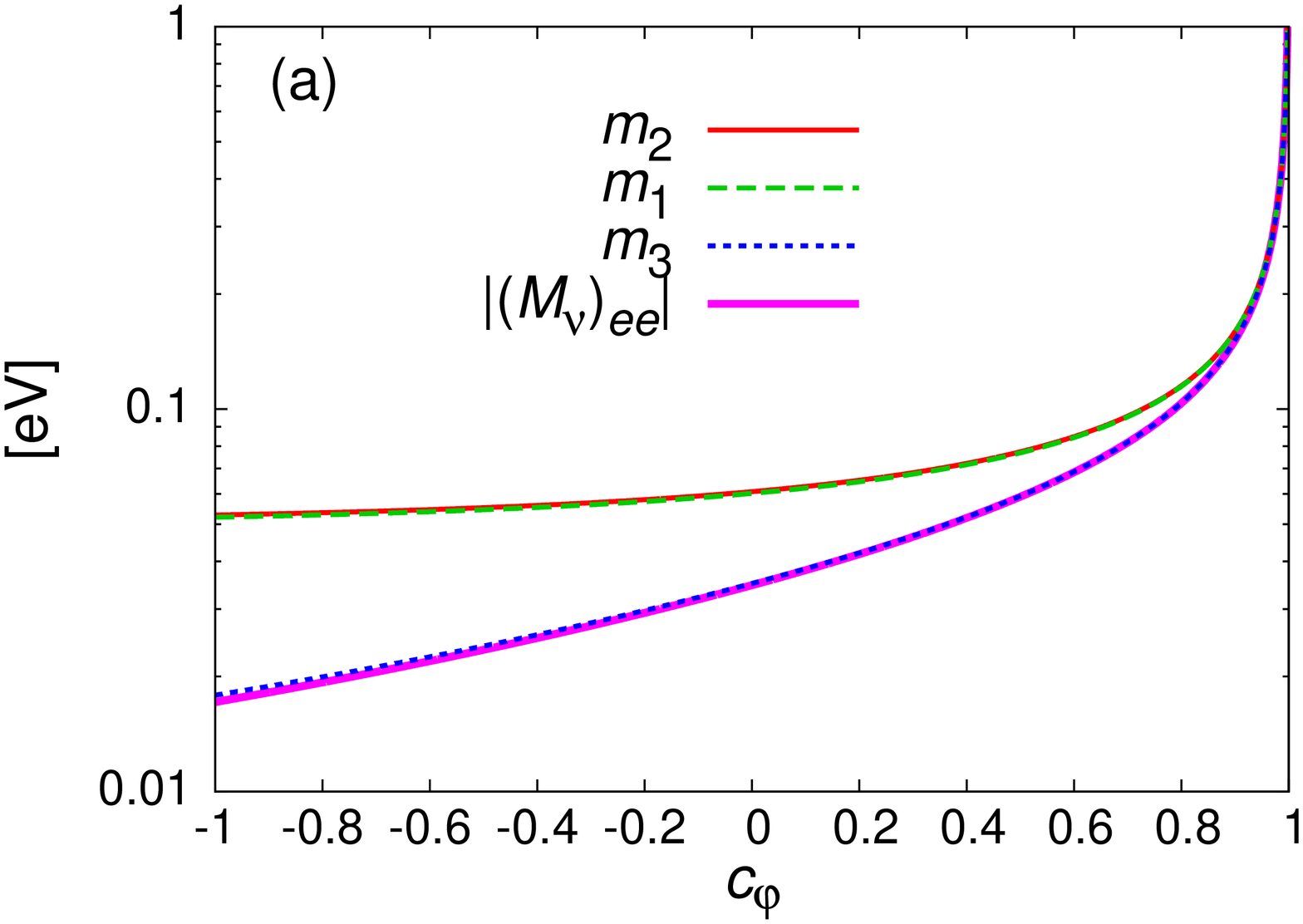} \
\includegraphics[scale = 0.3, origin = c, angle = -90]{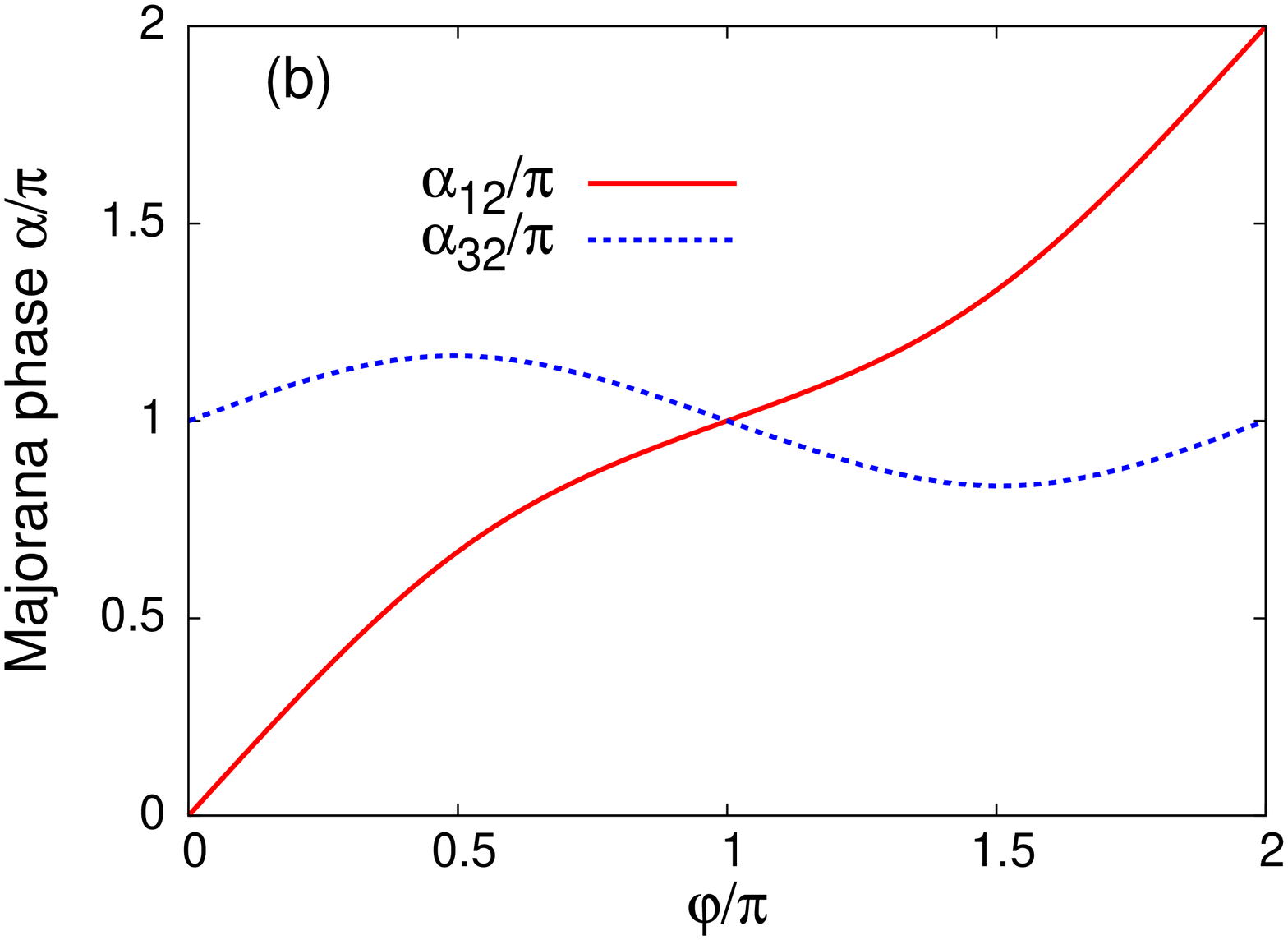}
\vspace*{-15mm}
\caption{
 (a) The $c_\varphi$ dependences of
$m_1$, $m_2$, $m_3$, and $|(M_\nu)_{ee}|$
are shown with
red solid, green dashed, blue dotted,
and bold solid magenta lines, respectively.
 Note that $m_2 \gtrsim m_1 > m_3 \gtrsim |(M_\nu)_{ee}|$.
 (b) The $\varphi$ dependences of two Majorana phases
$\alpha_{12}$ and $\alpha_{32}$ are shown
with red solid and blue dotted lines, respectively.
}
\label{fig:mass}
\end{center}
\end{figure}
 Neutrino masses in the A4ZM are expressed
with two mass parameters $|a|$ and $|b|$
and a phase $\varphi\equiv\text{arg}(b)$
as shown in Eqs.~(\ref{eq:m1})-(\ref{eq:m3}).
 It is clear that the A4ZM predicts a sum rule
\begin{eqnarray}
m_1 e^{i\alpha_{12}} + 2 m_2 + 3 m_3 e^{i\alpha_{32}} = 0 .
\label{eq:sumrule}
\end{eqnarray}
 Note that the simplest Zee model
predicts $m_1 + m_2 + m_3 = 0$
(no Majorana phases);
 for another example,
the simplest version of the Higgs Triplet Model
with softly broken $A_4$ symmetry
(A4HTM)~\cite{Fukuyama:2010mz}
shows $m_1 e^{i\alpha_{12}^{}} - 2m_2 - m_3 e^{i\alpha_{32}^{}} = 0$.
 Sum rules in other models are listed in \cite{Barry:2010yk}.
 Two mass parameters $|a|$ and $|b|$ are expressed as
\begin{eqnarray}
|a|^2
&=&
 \frac{ - \Delta m^2_{31} }{24 (1- c_\varphi^2)}
 \Bigl\{
  2 ( 3 + 2 r )
  + ( 1 - c_\varphi^2 ) ( 3 + 4 r )
\nonumber\\
&&\hspace*{30mm}
{}+ c_\varphi^{}
    \sqrt{ 4 ( 3 + 2 r )^2 - (1 - c_\varphi^2) ( 3 + 4 r )^2 }
 \Bigr\} ,
\label{eq:a}\\
|b|^2
&=&
 \frac{ - \Delta m^2_{31} }{24 (1- c_\varphi^2)}
 \Bigl\{
  2 ( 3 + 2 r )
  - ( 1 - c_\varphi^2 ) ( 3 + 4 r )
\label{eq:b}
\nonumber\\
&&\hspace*{30mm}
{}+ c_\varphi
    \sqrt{ 4 ( 3 + 2 r )^2 - (1 - c_\varphi^2) ( 3 + 4 r )^2 }
 \Bigr\} ,
\end{eqnarray}
where $c_\varphi^{}\equiv \cos\varphi$
and $r \equiv \Delta m^2_{21}/(-\Delta m^2_{31})$.
 We see that the A4ZM predicts $\Delta m^2_{31} < 0$
which is so-called inverted mass ordering.
 Lower bounds on $|a|$ and $|b|$ are given by
$c_\varphi = -1$.
 The lower bound on $|a|$ results in
\begin{eqnarray}
m_3^2
=
 4 |a|^2 - \Delta m^2_{21} + \Delta m^2_{31}
\geq
 (1.8\times 10^{-2}\,\eV)^2 .
\label{eq:bound_m3}
\end{eqnarray}
 The existence of the nontrivial lower bound on $m_3$
can be understood by the fact that
$m_3=0$ in Eqs.~(\ref{eq:m1})-(\ref{eq:m3})
conflicts with $\Delta m^2_{21} > 0$.
 Figure~\ref{fig:mass}(a) shows
behaviors of $m_i$ with respect to $c_\varphi$.
 The red thin solid, green dashed, and blue dotted lines
are for $m_1$, $m_2$, and $m_3$, respectively.

 The neutrinoless double beta decay
is the most promising phenomenon
of the lepton number violation
which is caused by Majorana neutrinos.
 The effective mass $|(M_\nu)_{ee}|$
which controls the decay
is given by
\begin{eqnarray}
| (M_\nu)_{ee} |^2
&=&
 4 |b|^2
=
 m_3^2 - \frac{\Delta m^2_{21}}{\,3\,}
\geq
 ( 1.7\times 10^{-2}\,\eV )^2 .
\label{eq:bound_mee}
\end{eqnarray}
 The $c_\varphi$ dependence of $|(M_\nu)_{ee}|$
is shown in Fig.~\ref{fig:mass}(a)
with the magenta bold solid line.
 Most of the region $|(M_\nu)_{ee}| \geq 1.7\times 10^{-2}\,\eV$
would be proved by the future experiments
(See \cite{Avignone:2007fu} for a review).
 Note that the simplest Zee model predicts $(M_\nu)_{ee}=0$;
 for another example,
the A4HTM gives a lower bound $| (M_\nu)_{ee} | \geq 0.0045\,\eV$
which allows rather smaller values than the expected sensitivities
in the future experiments.

 The $\varphi$ dependences of 
two Majorana phases ($\alpha_{12}$ and $\alpha_{32}$)
in Eqs.~(\ref{eq:m1}) and (\ref{eq:m3})
are shown in Fig.~\ref{fig:mass}(b).
 Red solid and blue dotted lines are used for
$\alpha_{12}^{}$ and $\alpha_{32}^{}$, respectively.
 We see that
$\alpha_{32}^{}$ is restricted as
$|\alpha_{32}^{} - \pi| \leq 0.2\pi$ in this model.

\subsection{Higgs boson decays into leptons}\label{sect:Hdecay}
\begin{table}[t]
\begin{tabular}{c||c|c|c|c}
 {}
 & \ $\BR(s^-_\alpha \to \ell\nu)$ \
 & \ $\mu \to e \bar{\nu}_\ell \nu_{\ell^\prime}^{}$ \
 & \ $\tau \to \ell \bar{\nu}_{\ell^\prime} \nu_{\ell^\dprime}^{}$ \
 & \ Matter effect, \
\\
 {}
 & \ $e\nu:\mu\nu:\tau\nu$
 & 
 &
 & $\nu e \to \nu e$
\\\hline\hline
 \ $s^\pm_\xi$ \
 & \ $0:1:1$
 & None
 & $\tau \to \mu \bar{\nu}_\mu \nu_\tau$
 & None
\\\hline
 \ $s^\pm_\eta$ \
 & \ $1:0:1$
 & None
 & $\tau \to e \bar{\nu}_e \nu_\tau$
 & $\varepsilon_{\tau\tau}^{eP_L}$
\\\hline
 \ $s^\pm_\zeta$ \
 & \ $1:1:0$
 & $\mu \to e \bar{\nu}_e \nu_\mu$
 & None
 & $\varepsilon_{\mu\mu}^{eP_L}$, ($\varepsilon_{\ell\ell}^{eP}$)
\end{tabular}
\caption{
 Phenomenological aspects of $s^\pm_\alpha$.
 The second column shows ratios of the leptonic decays
of each $s_\alpha^\pm$, where the flavors of neutrinos are summed up.
 The third and fourth column present
$\ell \to \ell^\prime \bar{\nu} \nu$
which can be affected by $s_\alpha^\pm$ mediations.
 The last column shows
contributions of $s_\alpha^\pm$ to effective four-Fermion couplings
which relate to the nonstandard matter effect for the neutrino oscillation.
 The indirect contribution to the effect
through the redefinition of $G_F$ is indicated with parentheses.
 See the main text for the definition of $\varepsilon_{\ell\ell^\prime}^{fP}$.
}
\label{tab:sp}
\end{table}

 The characteristic particles in the A4ZM are
$s^\pm_\alpha$ of a ${\bf 3}$ representation.
 The interactions of $s^+_\alpha$ with leptons
are given by
singlet Yukawa coupling matrices $F_\alpha$ in Eq.~(\ref{eq:Falpha}).
 The second column of Table~\ref{tab:sp} shows
the ratios of the branching ratios
of leptonic decays of $s_\alpha^\pm$.
 The flavors of neutrinos are summed up
because they will not be detected
at collider experiments.
 Leptonic decays of $s_\alpha^\pm$ have characteristic flavor structures
unless their masses degenerate
(e.g., for a large $\tan\beta$).
 Each of $s_\alpha^\pm$ has only two modes
as leptonic decays;
 for example,
$s_\xi^-$ decays into $\mu_L^{} \nu_{\tau L}^{}$
and $\tau_L^{} \nu_{\mu L}^{}$
with a common decay rate.
 Note that $s_\alpha^-$
can be easily distinguished
from $H_{Di}^-$
whose leptonic decays are dominated by
the decay into $\tau$.
 Therefore,
if some of $s_\alpha^\pm$ are light enough to be produced at the LHC
this model can be testable by measuring
leptonic decays of $s_\alpha^\pm$.

 The Yukawa couplings of $H_{Di}^+$
are shown in Eq.~(\ref{eq:Yi}).
 Since $\Phi_q$
and a combination $\Phi_\xi \equiv (\Phi_x + \Phi_y + \Phi_z)/\sqrt{3}$
have the $Z_3$-charge 1,
they behave as usual doublet scalar fields.
 The phenomenology of $\Phi_q$ and $\Phi_\xi$
(namely, $H_{D3}^\pm$, $G^\pm$, and neutral members)
is almost identical to the one in a type of the two-Higgs-doublet models,
which can be seen in
\cite{Barger:1989fj, Grossman:1994jb, Akeroyd:1994ga, Aoki:2009ha}.
 Other two-linear combinations
$\Phi_\eta
 \equiv (\Phi_x + \omega^2 \Phi_y + \omega \Phi_z)/\sqrt{3}$
and
$\Phi_\zeta
 \equiv (\Phi_x + \omega \Phi_y + \omega^2 \Phi_z)/\sqrt{3}$
have no vev and no contribution
to the mass matrix of charged leptons.
 They can cause flavor violations in their Yukawa interactions.
 Phenomenology of $\Phi_\eta$ and $\Phi_\zeta$
is the same as the one in a model discussed
in \cite{A4NR}
(See also \cite{Ma:2008ym}).
 Dominant leptonic decays of them are
$\Phi_\eta \to \overline{\tau_R} L_e$ and
$\Phi_\zeta \to \overline{\tau_R} L_\mu$.

\subsection{Lepton flavor violating decays of charged leptons}
\label{sect:LFV}
 The A4ZM does not give sizable BRs of $\meee$ and $\llg$
because they are forbidden by the remaining $Z_3$ symmetry;
 for example,
$\meg$ changes the $Z_3$-charge from $\omega$ (of $\mu$)
to $1$ (of $e$ and $\gamma$).
 The $Z_3$ symmetry allows only $\temm$ and $\tmee$ among six $\tlll$.
 Tree-level contributions to $\temm$ and $\tmee$
are dominated by
mediations of $(H_{D1}^0)^\dagger \equiv (\phi^0_\zeta)^\dagger$
and $(H_{D2}^0)^\dagger \equiv (\phi^0_\eta)^\dagger$, respectively%
\footnote
{
 Since $H_{D1}^0$ and $H_{D2}^0$ have $Z_3$-charges,
they can be dealt with by keeping them as complex scalars.
 In other words,
masses of $\text{Re}(H_{D1}^0)$ and $\text{Im}(H_{D1}^0)$
are the same.
}.
 The Yukawa couplings appear
as $m_\mu^2 m_\tau^2/v_3^4$ for the decay rate of $\teLmLmR$
while the rate of $\tmReLeL$ is governed by $m_e^2 m_\tau^2/v_3^4$~\cite{A4NR}.
 Therefore,
only $\teLmLmR$ can have a sizable decay rate in the A4ZM%
\footnote
{
 If both of
left-handed lepton doublets and
right-handed charged leptons
are made from ${\bf 3}$ representations,
$\BR(\tmee)$ can be also sizable as discussed in \cite{Ma:2009wi}.
 See also Appendix~\ref{app:R3}.
}.
 The signal of $\temm$ may exist just below
the current experimental limit,
$\BR(\temm)<1.7\times 10^{-8}$ at 90\,\%~CL~\cite{t3lBelle}
(see also \cite{Lees:2010ez}),
because constraints from
other LFV processes are satisfied automatically.
 The lack of LFV in the A4ZM is a good feature of the model
because the model can be excluded easily
by the searches of the LFV processes.
 The branching ratio for $\temm$ is given by
\begin{eqnarray}
\BR(\temm)
=
 \BR(\teLmLmR)
=
 \frac{m_\tau^2 m_\mu^2}{8 v_3^4 G_F^2 m_{H_{D1}^0}^4}
 \BR(\tau \to \mu \overline{\nu_\mu} \nu_\tau) ,
\end{eqnarray}
where $\BR(\tau \to \mu \overline{\nu_\mu} \nu_\tau) = 0.17$
and $\sqrt{2} G_F = 1/(v_3^2 + v_q^2)$.
 The bound $\BR(\temm)<1.7\times 10^{-8}$ results in
\begin{eqnarray}
m_{H_{D1}^0}^{}
> 17 \sqrt{1+\tan^2\beta}\,[\GeV] .
\end{eqnarray}
 Note that $m_{H_{D1}^0}^{}$ can not be $O(10)\,\GeV$
because the LEP bound for $e^+ e^- \to Z^\ast \to H_1 H_2$~\cite{Schael:2006cr}
($e^+ e^- \to Z^\ast \to (H_{D1}^0)^\dagger H_{D1}^0$ in our case)
results in $m_{H_{D1}^0}^{} \gtrsim 90\,\GeV$.
 If $\tan\beta$ is less than a several-times $10$,
$H_{D1}^0$ can be light enough to be discovered at the LHC\@.
 Even for $\tan\beta \gtrsim 100$,
$H_{D2}^0$ can be light.

 There is no remarkable constraint for $s_\alpha^\pm$
from the LFV decays of charged leptons
because $\llg$ are forbidden
as explained above.
 Thus,
$s_\alpha^\pm$ can be light
without caring about constraints from $\llg$.
 We mention that
the singlet scalar contribution
to the anomalous magnetic dipole moment of muon
has a minus sign%
\footnote
{
 The sign of the contribution
seems to be misunderstood as the plus sign
sometimes.
}
while a plus sign is favored
to explain experimental results
(See e.g., \cite{Teubner:2010ah} and references therein).
 Other exotic phenomena (lepton flavor conserving) of $s_\alpha^\pm$
are discussed below.

\subsection{Universality of $G_{\ell\ell^\prime}$}
\label{sect:GF}
 In the A4ZM,
$s_\alpha^\pm$ can contribute to $\ell \to \ell^\prime \bar{\nu} \nu$
as shown in the third and fourth columns of Table~\ref{tab:sp}.
 See \cite{Smirnov:1996bv} for the case with
the simplest Zee model.
 The effective coupling constants for $\ell \to \ell^\prime \bar{\nu} \nu$
are denoted as $G_{\ell\ell^\prime}$.
 Note that contributions of $s_\alpha^\pm$ are coherent
with the exchange of $W$ boson
and can give large effects in principle.
 Such coherent effects of doublet scalars
have the chirality suppression
because they couple with $\ell_R$.
 Contributions of $s_\xi^\pm$, $s_\eta^\pm$, and $s_\zeta^\pm$
to $G_{\ell\ell^\prime}$ are denoted as
$G_{\tau\mu}^{s_\xi^\pm}$,
$G_{\tau e}^{s_\eta^\pm}$,
and $G_{\mu e}^{s_\zeta^\pm}$,
respectively.
 Explicit forms of them are
\begin{eqnarray}
G_{\tau\mu}^{s_\xi^\pm}
\equiv
 \frac{ |(F_\xi)_{\tau\mu}|^2 }{ 4\sqrt{2}\,m_{s_\xi^\pm}^2 }
=
 \frac{f^2}{\sqrt{2}\,m_{s_\xi^\pm}^2} , \quad
G_{\tau e}^{s_\eta^\pm}
\equiv
 \frac{f^2}{\sqrt{2}\,m_{s_\eta^\pm}^2} , \quad
G_{\mu e}^{s_\zeta^\pm}
\equiv
 \frac{f^2}{\sqrt{2}\,m_{s_\zeta^\pm}^2} .
\end{eqnarray}
 Note that
$2 (\overline{\nu_\mu^c} P_L \tau) (\overline{\nu_\tau} P_R \mu^c)
= (\overline{\nu_\tau} \gamma^\mu P_L \tau)
  (\overline{\mu} \gamma_\mu P_L \nu_\mu)$.

 The Fermi coupling constant $G_F=1.17\times 10^{-5}\,\GeV^{-2}$ is
determined by $\mu \to e \overline{\nu} \nu$.
 Since $s_\zeta^\pm$ contributes to the decay,
the value of $G_F$ should be used for
$G_{\mu e} = G^W + G_{\mu e}^{s_\zeta^\pm}$
where $G^W\equiv g^2/(4\sqrt{2} m_W^2)$
is the contribution from $W$ boson.
 The extremely precise measurement of $\mu \to e \overline{\nu} \nu$ itself
does not mean an extremely stringent constraint
on $G_{\mu e}^{s_\zeta^\pm}$
although the interpretation of $G_F$ changes.
 Following \cite{Smirnov:1996bv}
where the exotic effect to the decay rate
($\propto G_{\mu e}^2$) was assumed to be
smaller than 0.1\% in order to avoid conflicting with
the electroweak precision tests,
we have
\begin{eqnarray}
|f|
<
 2.7\times 10^{-2}
 \left( \frac{m_{s_\zeta^\pm}^{}}{300\,\GeV} \right) .
\label{eq:Gme}
\end{eqnarray}

 On the other hand,
the contribution of $W$ is universal
for $\mu \to e\overline{\nu}\nu$,
$\tau \to e\overline{\nu}\nu$,
and $\tau \to \mu\overline{\nu}\nu$.
 Contributions of exotic particles may break the universality,
$G_{\mu e}=G_{\tau e}=G_{\tau\mu}$.
 Constraints from the test of the lepton universality of $G_{\ell\ell^\prime}$
(p.~549 of \cite{Nakamura:2010zzi})
can be written as
\begin{eqnarray}
\frac{G_{\tau\mu}}{G_F}
&=&
 1
 - \frac{G_{\mu e}^{s_\zeta^\pm}}{G_F}
 + \frac{G_{\tau\mu}^{s_\xi^\pm}}{G_F}
= 0.981 \pm 0.018 ,
\label{eq:Gtm}\\
\frac{G_{\tau e}}{G_F}
&=&
 1
 - \frac{G_{\mu e}^{s_\zeta^\pm}}{G_F}
 + \frac{G_{\tau e}^{s_\eta^\pm}}{G_F}
= 1.0012 \pm 0.0053 .
\label{eq:Gte}
\end{eqnarray}
 If we take a scenario of a large $\tan\beta$,
masses of $s_\alpha^\pm$ are almost degenerate
as $m_{s^\pm_{}}^{}\simeq m_{s_\alpha^\pm}^{}$.
 These constraints are then satisfied automatically
even for light $s_\alpha^\pm$
by virtue of the remaining $A_4$ symmetry in the Higgs sector.
 If $m_{s_\alpha^\pm}^{}$ do not degenerate,
$G_{\tau\mu}^{s_\xi^\pm}/G_F$ and $G_{\tau e}^{s_\eta^\pm}/G_F$
are constrained as $\lesssim O(0.01)$
by Eqs.~(\ref{eq:Gtm}) and (\ref{eq:Gte}).
 These constraints allow
$f = O(0.01)$ for $m_{s_\alpha^\pm}^{} = O(100)\,\GeV$.
 Therefore,
the A4ZM can be tested
if $s_\alpha^- \to \ell \nu$ are measured precisely at the LHC\@.

\subsection{Nonstandard interaction of neutrinos}
\label{sect:NSI}
 During the propagation of neutrinos in the ordinary matter,
the coherent forward scattering of them
on the matter ($e$, $u$, and $d$)
affects neutrino oscillations%
~\cite{Wolfenstein:1977ue,Mikheev:1986gs}.
 The so-called nonstandard interaction (NSI)
of neutrinos can give the nonstandard matter effect
on the neutrino oscillation%
~\cite{Wolfenstein:1977ue,NSI-matter}.
 The effective interaction for the exotic effect
is expressed conventionally as
\begin{eqnarray}
 2 \sqrt{2}\, G_F \varepsilon_{\ell\ell^\prime}^{fP}
 \left( \overline{f} \gamma^\mu P f \right)
 \left( \overline{\nu_\ell^{}} \gamma_\mu P_L \nu_{\ell^\prime}^{} \right),
\label{eq:NSI}
\end{eqnarray}
where $f=e, u ,d$ and $P= P_L, P_R$.
 Note that
$s_\alpha^\pm$ can contribute to the interaction
by using
$2 (\overline{\nu_{\ell^\prime}^c} P_L f)
   (\overline{f} P_R \nu_\ell^c)
= (\overline{f} \gamma^\mu P_L f)
  (\overline{\nu_\ell^{}} \gamma_\mu P_L \nu_{\ell^\prime}^{})$.
 The last column of Table~\ref{tab:sp}
shows possible $\varepsilon_{\ell\ell^\prime}^{fP}$
in the A4ZM\@.
 There is no contribution of $s_\xi^\pm$
to $\varepsilon_{\ell\ell^\prime}^{eP_L}$
because it does not couple with $e$.
 The contribution of $s_\eta^\pm$
to $\nu_\tau e \to \nu_\tau e$ is
given by $G_{\tau\mu}^{s_\eta^\pm}/G_F$.
 On the other hand,
$s_\zeta^\pm$ contributes directly
to $\nu_\mu e \to \nu_\mu e$
with $G_{\mu e}^{s_\zeta^\pm}/G_F$.
 In addition,
indirect contributions of $s_\zeta^\pm$
to $\varepsilon_{\ell\ell}^{eP}$ exist
through $G^W/G_F$ which is not the unity in this model
but rather $1-G_{\mu e}^{s_\zeta^\pm}/G_F$.
 We have
\begin{eqnarray}
\varepsilon_{ee}^{eP_L}
&=&
 0
 - (1+g_L^{e(SM)}) \frac{G_{\mu e}^{s_\zeta^\pm}}{G_F} ,
\label{eq:epee}\\
\varepsilon_{\mu\mu}^{eP_L}
&=&
 \frac{G_{\mu e}^{s_\zeta^\pm}}{G_F}
 - g_L^{e(SM)} \frac{G_{\mu e}^{s_\zeta^\pm}}{G_F} ,
\label{eq:epmm}\\
\varepsilon_{\tau\tau}^{eP_L}
&=&
 \frac{G_{\tau\mu}^{s_\eta^\pm}}{G_F}
 - g_L^{e(SM)} \frac{G_{\mu e}^{s_\zeta^\pm}}{G_F} ,
\label{eq:eptt}\\
\varepsilon_{\ell\ell}^{eP_R}
&=&
 0
 - g_R^{e(SM)} \frac{G_{\mu e}^{s_\zeta^\pm}}{G_F} ,
\label{eq:epR}
\end{eqnarray}
where $g_L^{e(SM)} = -0.269$
and $g_R^{e(SM)} = 0.231$.
 The first and second terms in the right hand-side
of Eqs.~(\ref{eq:epee})-(\ref{eq:epR}) correspond to
the direct and indirect contributions of $s_\alpha^\pm$,
respectively.
 See \cite{Davidson:2003ha} for model-independent constraints
on the NSI of neutrinos.
 In the A4ZM,
$\varepsilon_{\ell\ell^\prime}^{eP}$ are constrained
by Eqs.~(\ref{eq:Gme})-(\ref{eq:Gte})
from $\ell \to \ell^\prime \bar{\nu} \nu$.
 Values of $\varepsilon_{\ell\ell^\prime}^{eP}$
turn out unfortunately to be $\lesssim O(0.01)$
and smaller than the expected sensitivity~($\sim 0.1$)~\cite{NSI-nufact}
at the neutrino factory in the future.

\section{conclusions and discussions}\label{sect:concl}

 In this article,
we proposed the A4ZM
in which the softly broken $A_4$ symmetry
was introduced to the Zee model in the simplest way.
 The soft breaking term of $A_4$ is required
by the appropriate breaking pattern of $A_4$
to obtain the tribimaximal mixing
which agrees well with neutrino oscillation measurements.
 It was assumed that
the soft breaking of $A_4$ came from
the small breaking terms of the lepton number conservation.
 This assumption makes
an approximate $Z_3$ symmetry remain in this model.

 Realizing the tribimaximal mixing,
the A4ZM gives a sum rule
for mass eigenvalues of neutrinos,
$m_1 e^{i\alpha_{12}^{}} + 2m_2 + 3m_3 e^{i\alpha_{32}^{}} = 0$.
 The sum rule results in
$\Delta m^2_{31} < 0$ (the inverted mass ordering)
and gives the lower bound
$m_3 \geq 1.8\times 10^{-2}\,\eV$.
 The effective mass for the neutrinoless double beta decay
has a simple relation
$|(M_\nu)_{ee}|^2 = m_3^2 - \Delta m^2_{21}/3$
and the lower bound $|(M_\nu)_{ee}| \geq 1.7\times 10^{-2}\,\eV$.
 Since most of the region of the $|(M_\nu)_{ee}|$
will be probed in the future experiments,
this model presents a good prospect
of affirmative results in the experiments.

 The remaining $Z_3$ symmetry controls well the FCNC
which is necessary for the Zee model
to be consistent with neutrino oscillation data.
 Only $\BR(\temm)$
which is caused by the FCNC
can be sizable
among LFV decays of charged leptons in this model.
 This model will be excluded easily
if other LFV decays of charged leptons are discovered in the future.

 The characteristic particles in the A4ZM
are three $SU(2)_L$-singlet charged Higgs bosons
$s_\alpha^+$ ($\alpha=\xi,\eta,\zeta$)
which belong to a ${\bf 3}$ representation of $A_4$.
 We showed predictions about
the flavor structure of leptonic decays
$s_\alpha^- \to \ell \nu$;
for example,
$\BR(s_\xi^- \to \ell \nu)$
gives the ratios of the final states as
$e\nu:\mu\nu:\tau\nu = 0:1:1$.
 Since $\llg$ are almost forbidden in the A4ZM
by the remaining $Z_3$ symmetry,
$s_\alpha^\pm$ are not constrained stringently.
 Therefore,
it could be expected that some of $s_\alpha^\pm$ are
light enough to be produced at the LHC\@.
 Then,
the characteristic flavor structure of $\BR(s_\alpha^- \to \ell \nu)$
will allow this model to be explored.
 There are mild constraints
form $\ell \to \ell^\prime \bar{\nu} \nu$.
 The constraints are, however,
too strong to observe
nonstandard effects for neutrino oscillations in matter
with $\nu_\ell^{} e \to \nu_\ell^{} e$
in future experiments.

 Finally,
we mention some other models in which
neutrino masses are generated by loop diagrams.
 Also in the Ma model~\cite{Ma:2006km},
neutrino masses are given by one-loop diagrams
which are different from the diagrams used in the Zee model.
 A version of the model with the softly broken $A_4$ symmetry
was discussed in \cite{Ma:2008ym}.
 The two-loop and the three-loop diagrams are used for neutrino masses
in the Zee-Babu model~\cite{ZB}
and the Krauss-Nasri-Trodden model~\cite{Krauss:2002px,Cheung:2004xm},
respectively.
 If we try to introduce the $A_4$ symmetry to these two models,
there seems to be a difficulty;
 additional $SU(2)_L$-doublet scalar fields,
which will be introduced always by the $A_4$ symmetry,
allow that the neutrino masses in these models are generated
at the one-loop level identically to the Zee model.
 Another three-loop diagram is used
in the Aoki-Kanemura-Seto (AKS) model~\cite{Aoki:2008av}
which is compatible with multiple $SU(2)_L$-doublet scalars.
 The AKS model with the $A_4$ symmetry
will be discussed elsewhere~\cite{A4AKSM}.

\begin{acknowledgments}
 The work of T.F.\ is supported in part by the Grant-in-Aid for Science Research
from the Ministry of Education, Science and Culture of Japan
under Contracts No.~020540282 and 21104004.
 The work of K.T.\ is supported in part by the National Science Council of
Taiwan under Grant No.~NSC 99-2811-M-002-088.
\end{acknowledgments}

\appendix

\section{right-handed charged leptons of a ${\bf 3}$ representation}
\label{app:R3}

 In the main part of this article,
three right-handed charged leptons
($\psi_{1R}^-$, $\psi_{2R}^-$, and $\psi_{3R}^-$)
belong to three one-dimensional representations of $A_4$.
 Here,
we take another choice
that right-handed charged leptons are in a {\bf 3} representation.
 They are expressed as $\psi_{AR}^- (A=x,y,z)$.
 See also \cite{Ma:2009wi}
for a model with right-handed charged leptons 
of a {\bf 3} representation.
 The Yukawa terms of leptons with doublet scalar fields
are modified as
\begin{eqnarray}
{\mathcal L}_{\text{d-Yukawa}}
&=&
 y_q \bigl(\, \overline{\Psi_L}\, \psi_R \bigr)_{\bf 1} \Phi_q
 + y_s \bigl(
        \bigl(\, \overline{\Psi_L}\, \psi_R \bigr)_{{\bf 3}_s} \Phi
       \bigr)_{\bf 1}
 + y_a \bigl(
        \bigl(\, \overline{\Psi_L}\, \psi_R \bigr)_{{\bf 3}_a} \Phi
       \bigr)_{\bf 1}
 + \text{h.c.}
\end{eqnarray}
 With the vev alignment in Eq.~(\ref{eq:vev}),
flavor eigenstates of right-handed charged leptons are given by
\begin{eqnarray}
\begin{pmatrix}
 e_R\\
 \mu_R\\
 \tau_R
\end{pmatrix}
=
 U_R^\dagger
 \begin{pmatrix}
  \psi_{xR}^-\\
  \psi_{yR}^-\\
  \psi_{zR}^-
 \end{pmatrix} , \quad
U_R^\dagger
\equiv
 \frac{1}{\sqrt{3}}
 \begin{pmatrix}
  1 &        1 & 1\\
  1 & \omega^2 & \omega\\
  1 &   \omega & \omega^2
 \end{pmatrix} ,
\label{eq:right}
\end{eqnarray}
while left-handed leptons are still obtained by Eq.~(\ref{eq:lepton}).
 Their $Z_3$-charges are the same as before (Table~\ref{tab:Z3charge}).
 The masses of charged leptons are
\begin{eqnarray}
m_e
&=&
 \frac{1}{\sqrt{2}}\, v_q\, y_q
 + \sqrt{\frac{2}{\,3\,}}\, v_3^{}\, y_s ,
\\
m_\mu
&=&
 \frac{1}{\sqrt{2}}\, v_q\, y_q
 - \frac{1}{\sqrt{6}}\, v_3^{}\, y_s
 - i \frac{1}{\sqrt{2}}\, v_3^{}\, y_a ,
\\
m_\tau
&=&
 \frac{1}{\sqrt{2}}\, v_q\, y_q
 - \frac{1}{\sqrt{6}}\, v_3^{}\, y_s
 + i \frac{1}{\sqrt{2}}\, v_3^{}\, y_a .
\end{eqnarray}
 The Yukawa coupling constants can be expressed as
\begin{eqnarray}
y_q
&=&
 \frac{\sqrt{2}}{3 v_q} \left( m_e + m_\mu + m_\tau \right) ,
\\
y_s
&=&
 \frac{1}{\sqrt{6}\, v_3^{}} \left( 2 m_e - m_\mu - m_\tau \right) ,
\\
y_a
&=&
 \frac{i}{\sqrt{2}\, v_3^{}} \left( m_\mu - m_\tau \right) .
\end{eqnarray}
 The Yukawa matrices in Eq.~(\ref{eq:Yi})
are replaced by
\begin{eqnarray}
&&
Y_1
=
 \frac{\sqrt{2}}{3v_3^{}}
 \begin{pmatrix}
  0
   & - m_e - m_\mu + 2 m_\tau
   & 0\\
  0
   & 0
   & 2 m_e - m_\mu - m_\tau\\
  - m_e + 2 m_\mu - m_\tau
   & 0
   & 0
 \end{pmatrix} ,\nonumber\\
&&
Y_2
=
 \frac{\sqrt{2}}{3v_3^{}}
 \begin{pmatrix}
  0
   & 0
   & - m_e + 2m_\mu - m_\tau\\
  - m_e - m_\mu + 2 m_\tau
   & 0
   & 0\\
  0
   & 2 m_e - m_\mu - m_\tau
   & 0
 \end{pmatrix} ,\nonumber\\
&&
Y_3
=
 -s_\beta^{} Y_\xi
 + c_\beta^{} Y_q ,
\label{eq:YiR3}
\end{eqnarray}
where
\begin{eqnarray}
&&
Y_\xi
\equiv
 \frac{\sqrt{2}}{3v_3^{}}
 \begin{pmatrix}
  2 m_e - m_\mu - m_\tau
   & 0
   & 0\\
  0
   & - m_e + 2m_\mu - m_\tau
   & 0\\
  0
   & 0
   & - m_e - m_\mu + 2 m_\tau
 \end{pmatrix} ,
\\
&&
Y_q
\equiv
 \frac{\sqrt{2}}{3v_q}
 ( m_e + m_\mu + m_\tau )
 \begin{pmatrix}
  1 & 0 & 0\\
  0 & 1 & 0\\
  0 & 0 & 1
 \end{pmatrix} .
\end{eqnarray}
 Note that
$Y_G=c_\beta^{} Y_\xi+ s_\beta^{} Y_q$
does not change from Eq.~(\ref{eq:YG})
as expected.
 Although $M_\nu$ has more complicated form than Eq.~(\ref{eq:formMn}),
it can be divided into six parts (matrices)
which are proportional to $m_\ell^{} m_{\ell^\prime}^{}$.
 Each of these parts has no $\ell\ell^\prime$ element;
 for example,
a part proportional to $m_\mu m_\tau$
has zeros at $\mu\tau$ and $\tau\mu$ elements.
 Therefore,
$(M_\nu)_{\tau\tau}$ vanishes when $m_e$ and $m_\mu$ are ignored.
 Then,
we have again the sum rule in Eq.~(\ref{eq:sumrule})
for the tribimaximal mixing.
 Naive expectation on the size of $\theta_{13}$ in this case
will be $s_{13} \sim m_\mu/m_\tau \simeq 6\times 10^{-2}$
which is larger by 1 order of magnitude
than the value for the case in the main text.
 Results in Sec.~\ref{sect:pheno} do not change
except for the phenomenology of
doubletlike Higgs bosons.
 Table~\ref{tab:R3} shows ratios of $\BR(H_{Di}^- \to \ell \bar{\nu})$
where $m_e$ and $m_\mu$ are neglected for simplicity.
 Neutrino flavors are summed up.
 The second column shows results for the case
where right-handed charged leptons are of one-dimensional representations
while the third column is for those of a ${\bf 3}$ representation.
 Interactions of
$H_{D1}^0 \equiv (\phi_x^0 + \omega \phi_y^0 + \omega^2 \phi_z^0)/\sqrt{3}$
and
$H_{D2}^0 \equiv (\phi_x^0 + \omega^2 \phi_y^0 + \omega \phi_z^0)/\sqrt{3}$
with leptons are given also by $Y_1$ and $Y_2$ in Eq.~(\ref{eq:YiR3}),
respectively.
 These complex neutral scalars $H_{D1}^0$ and $H_{D2}^0$
are mass eigenstates
with $Z_3$-charges of $\omega^2$ and $\omega$, respectively.
 Note that $\tmee$ mediated by $H_{D1}^0$ and $(H_{D2}^0)^\dagger$
is not suppressed by $m_e/v_3$ and $m_\mu/v_3$ in this case.
 Thus
not only $\BR(\temm)$,
which is mediated by $(H_{D1}^0)^\dagger$ and $H_{D2}^0$,
but also $\BR(\tmee)$ can be sizable
as discussed in \cite{Ma:2009wi}.
 The current bound is
$\BR(\tmee)<1.5\times 10^{-8}$ at 90\,\%~CL~\cite{t3lBelle}
(see also \cite{Lees:2010ez}).
 The branching ratios are
\begin{eqnarray}
\BR(\tmee) = \BR(\temm)
&=&
 \frac{ m_\tau^4 }{162 v_3^4 G_F^2 }
 \left(
  \frac{1}{m_{H_{D1}^0}^4} + \frac{1}{m_{H_{D2}^0}^4}
 \right)
 \BR(\tau \to \mu \overline{\nu_\mu} \nu_\tau) ,
\end{eqnarray}
where $\BR(\tau \to \mu \overline{\nu_\mu} \nu_\tau) = 0.17$
and $\sqrt{2} G_F = 1/(v_3^2 + v_q^2)$.
 By using $\BR(\tmee)<1.5\times 10^{-8}$,
we have
\begin{eqnarray}
\left(
 \frac{1}{m_{H_{D1}^0}^4} + \frac{1}{m_{H_{D2}^0}^4}
\right)^{-\frac{1}{\,4\,}}
> 34 \sqrt{1+\tan^2\beta}\,[\GeV] .
\end{eqnarray}

\begin{table}[t]
\begin{tabular}{c||c|l}
 {}
 & \multicolumn{2}{|c}{$\BR(H_{Di}^- \to \ell \bar{\nu})$}
\\
 {}
 & \multicolumn{2}{|c}{$e\bar{\nu}:\mu\bar{\nu}:\tau\bar{\nu}$}
\\\hline
 $\ell_R$
 & \ ${\bf 1}$, ${\bf 1}^\prime$, ${\bf 1}^\dprime$ \
 & \hspace*{10mm} ${\bf 3}$
\\\hline\hline
 \ $H_{D1}^\pm$ \
 & \ $0:0:1$ \
 & \ \ $1:4:1$
\\\hline
 \ $H_{D2}^\pm$ \
 & \ $0:0:1$ \
 & \ \ $4:1:1$
\\\hline
 \ $H_{D3}^\pm$ \
 & \ $0:0:1$ \
 & \ \ $1:1:(1-3 s_\beta^2)^2$
\end{tabular}
\caption{
 Ratios of $\BR(H_{Di}^- \to \ell \bar{\nu})$
depending on representations of right-handed leptons.
 For simplicity
$m_e=m_\mu=0$ is used.
 Interactions of $G^\pm$ (NG boson) with charged leptons
are dominated by the interaction with $\tau$.
}
\label{tab:R3}
\end{table}


\end{document}